\documentstyle[prl,aps,epsf,rotate]{revtex}
\draft
\begin{document}
\twocolumn[\hsize\textwidth\columnwidth\hsize
     \csname @twocolumnfalse\endcsname

\title{Internal structures of electrons and photons and some
consequences in relativistic physics}

\author{W. A. Hofer} 
\address{Institut f\"ur Allgemeine Physik,
Technische Universit\"at Wien,
         A--1040 Vienna, Austria}

\maketitle


\begin{abstract}
The theoretical foundations of quantum mechanics and de Broglie--Bohm
mechanics are analyzed and it is shown that both theories employ a
formal approach to microphysics. By using a realistic approach
it can be established that the internal structures of particles comply
with a wave--equation. Including external potentials yields the
Schr\"odinger equation, which, in this context, is arbitrary due to 
internal energy components. The uncertainty relations are an expression 
of this, fundamental, arbitrariness. Electrons and photons can
be described by an identical formalism, providing formulations equivalent to
the Maxwell equations.  Electrostatic interactions justify the 
initial assumption of electron--wave stability: the stability of
electron waves can be referred to vanishing intrinsic fields of 
interaction. Aspect's experimental proof of non--locality is 
rejected, because these measurements imply a violation of the
uncertainty relations. The theory finally points out some
fundamental difficulties for a fully covariant formulation of
quantum electrodynamics, which seem to be related to the existing
infinity problems in this field. 
\end{abstract} 


\vskip2pc]

\section{Introduction}
It is commonly agreed upon that currently two fundamental frameworks
provide theoretical bases for a treatment of microphysical processes.
The standard quantum theory (QM) is accepted by the majority of the
physical community to yield the to date most appropriate account of
phenomena in this range. It can be seen as a consequence of the
Copenhagen interpretation, and its essential features are the following:
It (i) is probabilistic \cite{BOR26},
(ii) does not treat fundamental processes \cite{COP59,BOH52},
(iii) is restricted by a limit of description (uncertainty relations
\cite{HEI27}),
(iv) essentially non--local \cite{BEL64,ASP82},
and (v) related to classical mechanics \cite{PRI74}.

The drawbacks and logical inconsistencies of the theory have been
attacked by many authors, most notably by Einstein \cite{EPR35},
Schr\"odinger \cite{SCH92}, de Broglie \cite{BRO57}, Bohm \cite{BOH57},
and Bell \cite{BEL82}. Its application to quantum electrodynamics (QED)
provides an infinity problem, which has not yet been solved in any
satisfying way \cite{SCHW94}, for this reason QED has been critized 
by Dirac as essentially inadequate \cite{DIR84}.

These shortcomings have soon initiated the quest for an alternative
theory, the most promising result of this quest being the de
Broglie--Bohm approach to microphysics \cite{BRO26,BRO60,BOH52}. 
While the interpretations of
these two authors differ in detail, both approaches are centered around
the notion of ''hidden variables'' \cite{BOH52}. The de Broglie--Bohm
mechanics of quantum phenomena (DBQM) is still highly controversial
and rejected by most physicists, its main features are:
It (i) is deterministic \cite{BOH52},
(ii) has no limit of description (trajectories),
(iii) is highly non--local \cite{BEL66},
(iv) based on kinetic concepts,
and (v) ascribes a double meaning to the wave function
(causal origin of the quantum potentials and statistical
measure if all initial conditions are considered) \cite{HOL93}.

Analyzing the theoretical basis of these two theories, it is
found that QM and DBQM both rely on what could be called a
{\em formal approach} to microphysics. The Schr\"odinger equation
\cite{SCH26}

\begin{equation}
\left(-  \frac{\hbar^2}{2 m} \triangle + V \right) \psi = i \hbar 
\frac{\partial \psi}{\partial t}
\end{equation}

is accepted in both frameworks as a fundamental axiom; DBQM derives
the trajectories of particles from a quantum potential subsequent to
an interpretation of this equation, while the framework of standard
QM is constructed by employing, in addition, the Heisenberg commutation
relations \cite{HEI25}:

\begin{equation}
\left[X_{i}, P_{j} \right] = i \hbar \,\delta_{ij}
\end{equation}

Analogous forms of the commutation relations are used for second
quantization constitutional for the treatment of electromagnetic
fields in the extended framework of QED \cite{SCHW94}.

\section{Realistic approach}\label{approach}

Although these existing approaches have been highly successful in view
of their predictive power, they do not treat -- or only superficially
-- the question of the {\em physical} justification of their fundamental
axioms. This gap can be bridged by reconstructing the framework of
microphysics from a physical basis, and the most promising approach
seems to be basing it on the experimentally observed wave features of
single particles. For photons this point is trivial in view of wave
optics, while for electrons diffraction experiments by Davisson and
Germer \cite{DAV27} have established these wave features beyond doubt.

Using de Broglie's formulation of wave function properties \cite{BRO25}
and adapting it for a non--relativistic frame of reference we get 
(particle velocity $ |\vec u| << c_{0}$):

\begin{eqnarray}
\psi (x_{\mu}) = \psi_{0} \exp - i (k^{\mu} x_{\mu}) &
\longrightarrow & \psi (\vec r, t) = \psi_{0} \sin (\vec k \vec r
- \omega t) \nonumber \\
& \psi_{0} \in R &
\end{eqnarray}

The immediate consequence of the approach is a periodic and local
density of mass within the region occupied by the particle and which
is described by:

\begin{equation}
\rho (\vec r, t) = \rho_{0} \sin^2 (\vec k \vec r - \omega t)
\qquad \rho_{0} = C \psi_{0}^2
\end{equation}

The main reason that this approach -- leading to a local and
realistic picture of internal structures -- so far has remained
unconsidered is the dispersion relation of matter waves, based on
two independent statements \cite{BRO25,PLA01}. With the assumption
that energy of a free particle is kinetic energy of its inertial
mass $m$, the phase velocity of the matter wave is not equal to the
mechanical velocity of the particle, an obvious contradiction with the
energy principle: 

\begin{equation}
\lambda (\vec p) = \frac{h}{|\vec p|} \quad 
E (\omega) = \hbar \omega \longrightarrow
c_{phase} = 
\left(\frac{|\vec u|}{2}\right)_{non-rel}
\end{equation}

In case of real waves, however, the kinetic energy density is a
periodic function, and the energy principle then requires the
existence of an equally periodic intrinsic potential of particle
propagation. Using the total energy density for a particle of finite
dimensions and volume $V_{P}$ then yields the result, that internal
features described by a monochromatic plane wave of phase velocity
$c_{phase}$ comply with propagation of a particle with mechanical 
velocity $|\vec u|$:

\begin{eqnarray}
\frac{1}{2} \int_{V_{P}} dV \, \rho (\vec r, t=0) |\vec u|^2 = 
\frac{1}{2} \bar{\rho} V_{P} |\vec u|^2 = \frac{m}{2} |\vec u|^2 \nonumber
\\
W_{K} = \frac{m}{2} |\vec u|^2 \qquad 
W_{P} = \frac{m}{2} |\vec u|^2 \nonumber \\ 
W_{T} = W_{K} + W_{P} = m |\vec u|^2 =: \hbar \omega \\ 
c_{phase} = \lambda \nu = |\vec u| \nonumber
\end{eqnarray}

where $W_{K}$ denotes kinetic and $W_{P}$ potential energy of particle
propagation, the total energy is given by $W_{T}$.
The significance of intrinsic field components will be shown presently,
in this case the wave function and also the local density of mass 
comply with a wave equation:

\begin{eqnarray}
\triangle \psi (\vec r, t) - \frac{1}{|\vec u|^2} 
\frac{\partial^2 \psi (\vec r, t)}{\partial^2 t} = 0 \\
\triangle \rho (\vec r, t) - \frac{1}{|\vec u|^2} 
\frac{\partial^2 \rho (\vec r, t)}{\partial^2 t} = 0 
\nonumber 
\end{eqnarray}

\section{Quantum theory in a realistic approach}

For a periodic wave function $\psi (\vec r, t)$ the kinetic component
of particle energy can be expressed in terms of the Laplace operator
acting on $ \psi $, its value given by:

\begin{eqnarray}
|\vec u|^2 \triangle \psi = \frac{\partial^2 \psi}{\partial t^2} =
- \omega^2 \psi & \qquad &
W_{K} \psi = \frac{m}{2} |\vec u|^2 \psi = \frac{\hbar}{2} \omega \psi
\nonumber \\
\left(W_{K} + \frac{\hbar^2}{2 m} \right) \psi = 0 
& \longrightarrow &  W_{K} = - \frac{\hbar^2}{2 m} \triangle
\end{eqnarray}

If, furthermore, the total energy $W_{T}$ of the particle is equal to kinetic energy
and a local potential $ V(\vec r) $ the development of the wave
function is described by a time--independent Schr\"odinger equation
\cite{SCH26}:

\begin{equation}
\left(-  \frac{\hbar^2}{2m} \triangle + V(\vec r) \right) \psi (\vec r, t) =
W_{T} \psi (\vec r, t)
\end{equation}

Its interpretation in a realistic context is not trivial, though. If 
the volume of a particle is finite, then wavelengths and frequencies
become {\em intrinsic} variables of motion, which implies, due to
energy conservation, that the wave function itself is a measure for
the potential at an arbitrary location $ \vec r $. But in this case
the wave function must have physical relevance, and the EPR dilemma
\cite{EPR35} in this case will be fully confirmed. The result seems
to support Einstein's view, that quantum theory in this case cannot
be complete. The only alternative, which leaves QM intact, is the
assumption of zero particle volume: in the context of electrodynamics
this assumption leads to the,
equally awkward, result of infinite particle energy.

The second major result, that the Schr\"odinger equation in this
case is not an exact, but an essentially arbitrary equation, where
the arbitrariness is described by the uncertainty relations, can be
derived by transforming the equation into a moving reference frame
$ \vec r\,' = \vec r - \vec u t $:

\begin{equation}
\left(- \frac{\hbar^2}{2 m} \triangle' + V (\vec r\,' + \vec u t) \right)
\psi (\vec r\,') = W_{T} \psi (\vec r\,')
\end{equation}

Since the time variable in this case is undefined we may consider
the limits of the pertaining $k$--values of plane wave solutions
in one dimension. Resulting from a variation of
the potential $V (\vec r\,', t)$ they are given by:

\begin{eqnarray}
V_{1} = V (\vec r\,' + \vec u t_{1}) = V (\vec r\,') + \triangle V (t)
\nonumber \\
V_{0} = V (\vec r\,' + \vec u t_{0}) = V (\vec r\,') - \triangle V (t)
\\
\hbar \triangle k (t) = \frac{m \triangle V (t)}{\hbar k}
\nonumber 
\end{eqnarray}

If the variation results from the intrinsic potentials, which are
neglected in QM, then uncertainty of the applied potentials has as its
minimum the amplitude $\phi_{0}$, given by:

\begin{equation}
\phi_{0} = m u^2 = \triangle V
\end{equation}

Together with the uncertainty 
$ \triangle x = \frac{\lambda}{2} = 2 (x(\phi_{0}) - x(0)) $ for the
location, which is defined by the distance between two potential maxima
(the value can deviate in two directions), we get for the product:

\begin{equation}
\triangle x \triangle k \ge \frac{k \lambda}{2} \longrightarrow
\triangle x \triangle p \ge \frac{h}{2}
\end{equation}

Apart from a factor of $2 \pi$, which can be seen as a correction
arising from the Fourier transforms inherent to QM, the relation
is equal to the canonical formulation in quantum theory \cite{COH77}:

\begin{equation}
\triangle X_{i} \triangle P_{i} \ge \frac{\hbar}{2}
\end{equation}

The uncertainty therefore does not result from wave--features of particles,
as Heisenberg's initial interpretation suggested \cite{HEI27}, nor
is it an expression of a fundamental principle, as the Copenhagen
interpretation would have it \cite{COP59}. It is, on the contrary,
an {\em error margin} resulting from the fundamental assumption in
QM, i.e. the interpretation of particles as inertial mass aggregations.
Due to the somewhat wider frame of reference,
this result also seems to settle the long--standing controversy between
the {\em empirical} and the {\em axiomatic} interpretation of this
important relation:
although within the principles of QM the relation is an {\em axiom},
it is nonetheless a {\em result} of fundamental theoretical
shortcomings and not a physical principle.

Since this feature is inherent to any evaluation of the
Schr\"odinger equation, it also provides a reason for the
statistical ensembles pertaining to its solutions:
the structure of the ensembles treated in QM is based
on a fundamental arbitrariness of Schr\"odinger's equation,
which has to be considered in all calculations of measurement
processes \cite{HOF97B}.

\section{Classical electrodynamics}

The fundamental relations of classical electrodynamics (ED) are
commonly interpreted as axioms which cannot be derived from
physical principles \cite{JAC84}. However, within the realistic
approach they are an expression of the intrinsic features of
particle propagation and related to the proposed intrinsic
potentials, as can be shown as follows. We define the longitudinal
and intrinsic momentum $\vec p$ of a particle by:

\begin{equation}
\vec p (\vec r, t) := \rho (\vec r, t) \vec u \qquad
\vec u = constant
\end{equation}

From the wave equation and the continuity equation for $ \vec p $:

\begin{equation}\label{ed001}
\nabla^2 \vec p - \frac{1}{|\vec u|^2} 
\frac{\partial^2 \vec p}{\partial t^2} = 0  
\qquad \nabla \vec p + \frac{\partial \rho}{\partial t} = 0 
\end{equation}

the following expressions can be derived:

\begin{eqnarray}
\nabla^2 \vec p = - \nabla \frac{\partial \vec p}{\partial t}
- \nabla \times ( \nabla \times \vec p ) \nonumber \\
\frac{1}{|\vec u|^2} \frac{\partial^2 \vec p}{\partial t^2} =
\frac{\partial}{\partial t} \left( \frac{\bar{\sigma}}{|\vec u|^2}
\frac{1}{\bar{\sigma}} \frac{\partial \vec p}{\partial t} \right)
\end{eqnarray}

where $\bar{\sigma}$ shall be a dimensional constant to guarantee
compatibility with electromagnetic units. With the definition of
electromagnetic $\vec E$ and $\vec B$ fields by:

\begin{eqnarray}\label{ed002}
\vec E (\vec r, t) & := & - \nabla \frac{1}{\bar{\sigma}} 
\phi (\vec r, t) + \frac{1}{\bar{\sigma}} 
\frac{\partial \vec p}{\partial t} \nonumber \\
\vec B (\vec r, t) & := & - \frac{1}{\bar{\sigma}}
\nabla \times \vec p
\end{eqnarray}

where $ \phi (\vec r, t)$ shall denote some electromagnetic potential,
we derive the following expression:

\begin{eqnarray}
\frac{\partial}{\partial t} \nabla \left(
\frac{1}{|\vec u|^2} \phi + \rho \right) + \bar{\sigma}
\left( \frac{1}{|\vec u|^2}
\frac{\partial \vec E}{\partial t} - \nabla \times \vec B
\right) = 0
\end{eqnarray}

Since the total intrinsic energy density 
$ \phi_{T} = \rho |\vec u|^2 + \phi $ is, for a single particle
in constant velocity, an intrinsic constant, the following
equation is valid for every micro volume:

\begin{eqnarray}\label{ed003}
\frac{1}{|\vec u|^2}
\frac{\partial \vec E}{\partial t} = \nabla \times \vec B
\end{eqnarray}

From the definition of electromagnetic fields we get, in addition,
the following expression:

\begin{eqnarray}
\nabla \times \vec E = - \frac{\partial \vec B}{\partial t} 
\end{eqnarray}

which equals one of Maxwell's equations. Including the source
equations of ED and the definitions of current density $\vec J$
as well as the magnetic $\vec H$ field:

\begin{eqnarray}\label{ed004}
\nabla (\epsilon \vec E) = \nabla \vec D = \sigma \qquad
\nabla \vec B = 0 \\
\nabla \vec J (\vec r,t) := - \dot{\sigma} \qquad
\vec H := \mu^{-1} \vec B
\end{eqnarray}

where $\mu$ and $\epsilon$ are the permeability and the dielectric
constant, which are supposed invariant in the micro volume, and computing
the source of (\ref{ed003}) and the time derivative of (\ref{ed004}), 
we get for the sum:

\begin{eqnarray}
\nabla \left( -  \frac{1}{|\vec u|^2}
\frac{\mu^{-1} \partial \vec E}{\partial t} + \vec J
+ \frac{\partial \vec D}{\partial t} \right) = 0
\end{eqnarray}

For a constant of integration equal to zero we then obtain the
inhomogeneous Maxwell equation \cite{JAC84}:

\begin{eqnarray}
\vec J + \frac{\partial \vec D}{\partial t} =
\nabla \times \vec H
\end{eqnarray}
 
The energy principle of material waves can be identified as 
the classical Lorentz condition. To this aim we use the continuity
equation (\ref{ed001}) and the definition of electric fields
(\ref{ed002}). For linear 
$\nabla \times \nabla \times \vec p = 0$ and uniform motion
$ \bar{\sigma} \vec E = 0$ the equations lead to:

\begin{eqnarray}
\frac{1}{|\vec u|^2} \frac{\partial \phi}{\partial t}
- \nabla \vec p = 0
\end{eqnarray} 

In case of $|\vec u| = c_{0}$, a
comparison with the classical Lorentz condition \cite{JAC84}:

\begin{eqnarray}
\nabla \vec A + \frac{1}{c_{0}}
\frac{\partial \phi}{\partial t} = 0
\end{eqnarray}

yields the result, that the vector potential of ED is
related to the intrinsic momentum of the particle:

\begin{eqnarray}\label{ed005}
\vec A = - c_{0} \vec p = \frac{1}{\alpha} \vec p
\end{eqnarray}

It can thus be said that the Maxwell equations follow from 
intrinsic properties of particles in constant motion. But the 
result equally means, that electrodynamics and quantum theory 
{\em must have} the same dimensional level: both theories can be
seen as a {\em limited account} of intrinsic particle properties.

\section{Internal structures of photons and electrons}

On this basis a mathematical description of internal properties
of photons and electrons can be given, which comprises, apart
from the kinetic and longitudinal properties, also the transversal
features of electrodynamics. The two aspects of single particles
are related by the proposed intrinsic potential 
(see Section \ref{approach}), in case of photons it leads to
a total intrinsic energy density described by Einstein's
energy relation \cite{EIN06}. To derive the intrinsic properties
of photons we proceed from the wave equation (\ref{ed001}), the
energy relation (\ref{ed005}) and the equation for the change of
the vector potential in ED:

\begin{eqnarray}
\frac{1}{c_{0}} \frac{\partial \vec A}{\partial t} =
- \nabla \phi - \vec E
\end{eqnarray}

A wave packet shall consist of an arbitrary number $N$ of 
monochromatic plane waves:

\begin{eqnarray}
\vec p & = & \sum\limits_{i=1}^{N} p_{i}^0 {\vec e}^{k_{i}}
\sin^2 \left(\vec k_{i} \vec r - \omega_{i} t \right) \nonumber \\
\phi & = & \sum\limits_{i=1}^{N} \phi_{i}^0
\cos^2 \left(\vec k_{i} \vec r - \omega_{i} t \right)
\end{eqnarray}

where $p_{i}^0$ is the amplitude of longitudinal momentum, and
$\phi_{i}^0$ the corresponding intrinsic potential of a component
$i$. For a single
component of this wave packet, which shall define the {\em photon},
we get the equation:

\begin{eqnarray}
p_{i}^0 k_{i} - \frac{\omega_{i}}{c_{0}^2} \phi_{i}^0 = 0
\end{eqnarray}

Together with the relation $p_{i}^0 = \rho_{ph,i}^0 c_{0}$ for 
the momentum and the dispersion of plane waves in a vacuum 
$ c_{0} = \omega_{i}/k_{i}$ it leads to the following
expression:

\begin{eqnarray}
\phi_{i}^0 =  \rho_{ph,i}^0 c_{0}^2 \frac{c_{0} k_{i}}{\omega_{i}}
= \rho_{ph,i}^0 c_{0}^2 
\end{eqnarray}

The intrinsic energy of photons thus complies with Einstein's 
energy relation.  With the expression for the potential $\phi$
in classical electrodynamics:

\begin{eqnarray}
\phi_{em} = \frac{1}{8 \pi} \left(
\vec E^2 + \vec B^2 \right)
\end{eqnarray}

the electromagnetic and transversal fields for a photon are
given by:

\begin{eqnarray}
\vec E_{i} & = & \vec e^{t} c_{0} \sqrt{4 \pi \rho_{ph,i}^0}
\cos \left(\vec k_{i} \vec r - \omega_{i} t \right) \nonumber \\
\vec B_{i} & = & \left(\vec e^{k} \times \vec e^{t} \right)
c_{0} \sqrt{4 \pi \rho_{ph,i}^0}
\cos \left(\vec k_{i} \vec r - \omega_{i} t \right)
\end{eqnarray}

The units of electromagnetic fields are in this case dynamical
units, they can be in principle determined from the energy 
relations of electromagnetic fields and their relation to the
intrinsic momentum of a photon. These electromagnetic fields
are, furthermore, {\em causal} and not statistical variables
resulting from the intrinsic potentials of single photons. This
result, essentially incompatible with the current framework of
QED \cite{SCHW94}, has been derived by Renninger on the basis
of a gedankenexperiment \cite{REN53}.

For electrons the same procedure and a velocity $ \vec u < c_{0}$
leads to similar expressions of longitudinal and transversal
properties.

\begin{eqnarray}
\vec p & = &  \rho_{el}^0 {\vec u}
\sin^2 \left(\vec k \vec r - \omega t \right) \nonumber \\
\phi_{em} & = & \rho_{el}^0
\cos^2 \left(\vec k \vec r - \omega t \right) \nonumber \\
\phi_{el}^0 & = & \phi_{em} + \rho_{el} |\vec u|^2 =
\rho_{el}^0 |\vec u|^2 = const  \\
\nonumber \\
\vec E_{el} & = & \vec e^{t} |\vec u| \sqrt{4 \pi \rho_{el}^0}
\cos \left(\vec k \vec r - \omega t \right) \nonumber \\
\vec B_{el} & = & \left(\vec e^{k} \times \vec e^{t} \right)
|\vec u| \sqrt{4 \pi \rho_{el}^0}
\cos \left(\vec k \vec r - \omega t \right) \nonumber
\end{eqnarray}

The intrinsic electromagnetic fields are of transversal polarization
and solutions of the Maxwell equations for $ |\vec u| < c_{0}$
(subluminal solutions). These theoretical results establish also, that 
electrodynamics and quantum theory are not only formally analogous,
a result frequently used in interference models, but that they are
complementary theories of single particle properties.

\section{Spin in quantum theory}

One of the most difficult concepts in quantum theory is the
property of particle spin \cite{UHL25}. This is partly due to 
its abstract features, partly also to its relation with the 
magnetic moment, in electrodynamics a well defined 
vector of a defined orientation in space. To relate particle
spin to the polarizations of intrinsic fields, we first have to
consider the definition of the magnetic field (\ref{ed002})
not only for the intrinsic but, in case of electrons, also for
the external magnetic fields due to curvilinear motion. As can
be derived from the solution for $\vec u$ in a homogeneous
magnetic field $\vec B = B_{0} \vec e^{z}$, the definition 
in this case has to be modified for electrons by a factor of two, it
then reads:

\begin{eqnarray}
\vec B_{el} = - \frac{1}{2 \bar{\sigma}} \nabla \times
\vec p_{el} = \vec B_{ext}
\end{eqnarray} 

where the subscript denotes its relation to external fields.
The justification of this modification is the fact, that the 
intrinsic magnetic fields of electrons cannot be derived in 
quantum theory due to its fundamental assumptions.
On this basis particle spin of photons and electrons can be 
deduced from the expression for energy of magnetic interactions
in ED:

\begin{eqnarray}
W = - \mu \vec B
\end{eqnarray}

and by four assumptions: (i) The energy of electrons or photons
is equal to the energy in quantum theory, (ii) the magnetic field
referred to is the intrinsic (photon) or external (electron)
magnetic field, (iii) the frequency $\omega$ can be interpreted
as a frequency of rotation, and (iv) the magnetic moment is described 
by the relation in quantum theory:

\begin{eqnarray}
\vec \mu = g_{s} \frac{e}{2 m c_{0}} \vec s
\end{eqnarray}

For a photon the energy is the total energy, and with:

\begin{eqnarray}
W_{ph} = \hbar \omega \qquad \vec B_{ph} = 
- \frac{1}{\bar{\sigma}} \nabla \times \vec p \approx 
\frac{2 \bar{\rho}}{\bar{\sigma}} \vec \omega
\end{eqnarray}

as well as the ratio $ \bar{\rho}/\bar{\sigma} = m/e$,
and considering, in addition, that the magnetic fields in the
current framework are $c_{0}$ times the magnetic fields
in classical electrodynamics, we obtain the relation:

\begin{eqnarray}
W_{ph} = \hbar \omega = g_{ph} \vec{\omega} \vec s_{ph}
\end{eqnarray}

For constant $g_{ph}$ the relation can only hold, if:

\begin{eqnarray}
\vec s_{ph} \quad || \quad \vec \omega \longrightarrow
\vec s_{ph} \vec {\omega} = s_{ph} \omega \qquad
g_{ph} s_{ph} = \hbar 
\end{eqnarray}

The energy relation of electrodynamics is only consistent with
the energy of photons in quantum theory if they possess an
intrinsic spin of $\hbar$, a gyromagnetic ratio of 1, and if
the direction of spin--polarization is equal to the direction
of the intrinsic magnetic fields.

The same calculation for electrons, where the energy and the
magnetic fields are given by (intrinsic potentials of electrons
remain unconsidered in QM):

\begin{eqnarray}
W_{el} = \frac{1}{2} \hbar \omega \qquad \vec B_{el} = 
- \frac{1}{2 \bar{\sigma}} \nabla \times \vec p \approx 
\frac{\bar{\rho}}{\bar{\sigma}} \vec \omega                               
\end{eqnarray}

leads to the same result for the product of $s_{el}$ and 
$g_{el}$, namely:

\begin{eqnarray}
\vec s_{el} \quad || \quad \vec \omega \longrightarrow
\vec s_{el} \vec {\omega} = s_{el} \omega \qquad
g_{el} s_{el} = \hbar 
\end{eqnarray}
 
The difference to photons ($ g_{el} = 2$ and $s_{el} = \hbar/2$)
seems to originate from the assumptions of Goudsmit and Uhlenbeck
\cite{UHL25}, that the energy splitting due to the electron spin
in hydrogen atoms should be symmetric to the original state. This
general multiplicity of two for the spin states allows only for
the given solution. The direction of spin polarizations is also 
for electrons equal to the polarization of intrinsic fields.

\section{Bell's inequalities and Aspect's measurements}

We may reconsider Aspect's experimental proof of non--locality
\cite{ASP81}, based on Bell's inequalities \cite{BEL64} from the
viewpoint of the derived quality of photon spin. Adopting the
view on spin--conservation of quantum theory, the measurement
must account for the spin correlations of two photons with an
arbitrary angle $\vartheta$ of polarization in the states +1 
and -1, respectively.

Since the intrinsic magnetic fields oscillate with
\mbox{$\vec B = \vec B_{0} \cos (\vec k \vec r - \omega t)$}, the
spin variable $s(x)$ {\em cannot} remain constant, but must
equally oscillate from -1 to +1. And in this case a valid
measurement of spin polarizations $s(x)$ of the two particles is
only possible, if the variable is measured in the interval
$\triangle t < \tau/2$. The local resolution of the measuring device
must therefore be equal to $\triangle x < \lambda/2$. 

But as demonstrated in the deduction of the uncertainty relations,
this interval is lower than the local uncertainty in quantum
theory. A valid measurement of spin correlations thus violates
the uncertainty relation and, for this reason, {\em cannot} be
interpreted within the limits of quantum theory, which means,
evidently, that it cannot be interpreted with Bell's inequalities.
From this viewpoint Aspect's measurements are unsuitable for
a proof of non--locality, they can therefore not contradict
the current framework, which is essentially a local one.

\section{Electron photon interactions}

Defining the Lagrangian density by total energy density of an electron
in an external field $\phi$, including total energy of a presumed photon,
we may state:

\begin{eqnarray}
        {\cal  L} \, := \, T - V =
        \rho_{el}^{0} {{\dot{x}}_{i}^2} + \rho_{ph}^{0} c^2 -
        \sigma_{el}^{0} \phi
\end{eqnarray}

\begin{eqnarray*}
        \rho_{el}^{0} &=& constant \quad
        \dot{x}_{i}^2 = \dot{x}_{i}^2 (\dot{x}_{i})\\
        \rho_{ph}^{0} &=& \rho_{ph}^{0} (\dot{x}_{i})\quad
        \phi = \phi (x_{i})
\end{eqnarray*}

where $ \rho_{el}^{0} $ is the amplitude of electron density,
$ \rho_{ph}^{0} $ the amplitude of photon density, and $\sigma_{0}$
the amplitude of electron charge.
An infintesimal variation with fixed endpoints yields the result:

\begin{equation}\label{ep020}
        \int\limits_{t_{1}}^{t_{2}} dt \int d^3 x \left\{
        \sigma_{el}^{0} \,\frac{\partial \phi}{\partial x_{i}}
        + \frac{d}{d t} \left( \rho_{el}^{0}
        \frac{\partial \dot{x}_{i}^2}{\partial \dot{x}_{i}}
        + c^2 \frac{\partial \rho_{ph}^{0}}{\partial \dot{x}_{i}}
        \right)\right\} \delta x_{i}  = 0
\end{equation}

Therefore the following expression is valid:

\begin{equation}\label{ep021}
        \rho_{ph}^{0}c^2 = - \int dt \underbrace{
        \sigma_{el}^{0} \dot{x}_{i} \nabla \phi
        }_{- {\vec j}_{0} {\vec E}  =
        -  \partial V_{em}/\partial t }
        - \rho_{el}^{0} \dot{x}_{i}^2
        = \,+ \,V_{em} - \rho_{el}^{0} \dot{x}_{i}^2
\end{equation}

where we have used a relation of classical electrodynamics. From
(\ref{ep021}) and (\ref{ep020}) it can be derived that the partial
differential of ${\cal  L}$ may be written:

\begin{eqnarray}
        \frac{\partial {\cal L}}{\partial \dot{x}_{i}} =
        \frac{\partial V_{em}}{\partial \dot{x}_{i}}
\end{eqnarray}

Since the dependency of photon density on the velocity of the
electron is unknown, we may consider a small variation of electron
velocity and evaluate, with (\ref{ep021}), the Hamiltonian of the
system in first order approximation of a Taylor series. In this
case we get:

\begin{equation}\label{ep024}
        H = \frac{\partial {\cal L}}{\partial \dot{x}_{i}} \dot{x}_{i}
        - L \approx V_{em} - {\cal L} = \sigma_{el}^0 \enspace \phi
\end{equation}

The result seems paradoxical in view of kinetic energy of the
moving electron, which does not enter into the Hamiltonian.
Assuming, that an inertial particle is accelerated in an
external field, its energy density after interaction with
this field would only be altered according to its alteration of
location. The contradiction with the energy principle is
only superficial, though. Since the particle will have been
accelerated, its energy density {\em must} be changed. If
this change does not affect its Hamiltonian, the only
possible conclusion is, that photon energy has equally
been changed, and that the energy acquired by acceleration
has simultaneously been emitted by photon emission.
The initial system was therefore over--determined, and
the simultaneous existence of an external field {\em and}
interaction photons is no physical solution to the
interaction problem.

The process of electron
acceleration then has to be interpreted as a process of simultaneous
photon emission: the acquired kinetic energy is balanced
by photon radiation.
A different way to describe the same result,
would be saying that electrostatic interactions are
accomplished by an exchange of photons: the potential of
electrostatic fields then is not so much a function of
location than a history of interactions.
This can be shown by calculating the Hamiltonian of electron
photon interaction:

\begin{eqnarray}
H_{0} = \rho_{el}^{0}\, \dot{x}_{i}^2 + \sigma_{el}^{0} \phi
\qquad
H = \sigma_{el}^{0}\, \phi \nonumber\\
H_{w} = H - H_{0} = - \rho_{el}^{0} \,\dot{x}_{i}^2 =
\rho_{ph}^{0} \,c_{0}^2
\label{ep025}
\end{eqnarray}

But if electrostatic interactions can be referred to an exchange of
photons, and if these interactions apply to accelerated electrons,
then an electron in constant motion does not possess an intrinsic
energy component due to its electric charge: electrons in constant
motion are therefore stable structures.

It is evident that neither emission, nor absorption of photons at this stage
does show discrete energy levels: the alteration of velocity can be
chosen arbitrarily small.
The interesting question seems to be, how quantization fits
into this picture of continuous processes.
To understand this paradoxon, we consider the
transfer of energy due to an infinitesimal interaction
process. The differential of energy is given by:

\begin{equation}
        d\,W = A\cdot\,dt\,\rho_{ph}^{0}\,c^2
\end{equation}

A denotes the cross section of interaction. Setting $ dt $
equal to one period $ \tau $, and considering, that the
energy transfer during this interval will be $ \alpha\,W $,
we get for the transfer rate:

\begin{equation}
        \alpha\,\frac{d\,W}{d t} =
        \frac{A\,\lambda\,\rho_{ph}^{0}\,c^2}
        {d t} = \alpha\cdot\frac{V_{ph}\,\rho_{ph}^{0}\,c^2}{\tau} =
        \alpha\, \frac{h\,\nu}{\tau}
\end{equation}

Since total energy density of the particle as well as the
photon remains constant, the statement is generally valid.
And therefore the transfer rate in interaction processes will
be:

\begin{equation}
        \frac{d\,W}{d t} = \frac{d}{d t}\,m\,{\vec   u}^2 =
        \hbar\cdot\frac{\omega}{\tau} = h\cdot\nu^2
\end{equation}

The term energy quantum is, from this point of view, not quite
appropriate. The total value of energy transferred depends, in
this context, on the region subject to interaction processes, and
equally on the duration of the
emission process. The view taken in quantum theory is therefore
only a good approximation: a thorougher concept, of which
this theory in its present form is only the outline,
will have to account for {\em every} possible variation in the
interaction processes.

But even on this, limited, basis of understanding,
Planck's constant, commonly considered the fundamental value
of energy  quantization  is the fundamental constant not of
energy values, but of transfer rates in dynamic processes.
Constant transfer rates furthermore have the consequence, that
volume and mass values of photons or electrons become
irrelevant: the basic relations for energy and dispersion
remain valid regardless of actual quantities.
Quantization then is, in short, a {\em result of energy transfer} and
its characteristics.

\section{Consequences in relativistic physics}

The framework developed has some interesting consequences in 
relativistic physics. So far all expressions derived are only valid
in a non--relativistic inertial frame. To estimate its impact on
Special Relativity we may consider the wave equation in a moving frame
of reference, and estimate the result of energy measurments in the
system in motion $S'$ and the system at rest $S$. From the wave equation
in $S' = S'(V)$:

\begin{eqnarray}\label{ed035}
\triangle' p_{x}' - \frac{1}{u_{x}'^2}
\frac{\partial^2 p_{x}'}{\partial t'^2} = 0 \qquad
p_{x}' = \rho' u_{x}'
\end{eqnarray}

With the standard Lorentz transformation:

\begin{eqnarray}\label{ed036}
x'^{\mu} = \Lambda^{\mu}_{\nu} x^{\nu} \qquad
\Lambda^{\mu}_{\nu} = \left(
\begin{array}{rrrr}
\gamma & - \beta \gamma & 0 & 0 \\
- \beta \gamma & \gamma & 0 & 0 \\
0 & 0 & 1 & 0 \\
0 & 0 & 0 & 1
\end{array}
\right)
\end{eqnarray}

The differentials in $S'$ are given by:

\begin{eqnarray}\label{ed037}
\triangle' =
\left(\frac{\partial x}{\partial x'}  \right)^2 \triangle =
(1 - \beta^2) \triangle        \\
\frac{\partial^2}{\partial t'^2} =
\left(\frac{\partial t}{\partial t'}  \right)^2
\frac{\partial^2}{\partial t^2}  = (1 - \beta^2)
\frac{\partial^2}{\partial t^2} \nonumber
\end{eqnarray}

Density of mass and velocity in $S'$ are described by
(transformation of velocities according to the
Lorentz transformation, $\alpha$ presently undefined):

\begin{eqnarray}\label{ed038}
\rho' = \alpha \cdot \rho \qquad
u_{x}' = \frac{u_{x} - V}{1 - \frac{u_{x} V}{c^2}}
\end{eqnarray}

Then the transformation of the wave equations yields the
following wave equation for density $\rho$:

\begin{eqnarray}\label{ed039}
\triangle \rho - \frac{1}{u_{x}'^2}
\frac{\partial^2 \rho}{\partial t^2} = 0
\end{eqnarray}

Energy measurements in the system $S$ at rest must be measurements
of the phase--velocity $c_{ph}$ of the particle waves. Since,
according to the transformation, the measurement in $S$ yields:

\begin{eqnarray}\label{ed040}
c_{ph}^2 (S) = u_{x}'^2
\end{eqnarray}

the total potential $\phi_{0}$, measured in $S$ will be:

\begin{eqnarray}\label{ed041}
\frac{\phi_{0}(S)}{\rho_{S}} = u_{x}'^2  \quad \Rightarrow \quad
\phi_{0} (S) = \frac{\rho_{S}}{\rho_{S'}} \,\phi_{0} (S')
\end{eqnarray}

And if density $\rho_{S'}$ transforms according to a relativistic
mass--effect with:

\begin{eqnarray}\label{ed042}
\rho_{S'} = \gamma \, \rho_{S}
\end{eqnarray}

then the total intrinsic potential measured in the system $S'$ will
be higher than the potential measured in $S$:

\begin{eqnarray}\label{ed043}
\phi_{0} (S') = \gamma \phi_{0} (S)
\end{eqnarray}

But the total intrinsic energy of a particle with volume $V_{P}$
will not be affected, since the x--coordinate will be, again
according to the Lorentz transformation, contracted:

\begin{eqnarray}\label{ed044}
V_{P} (S') &=& \frac{1}{\gamma}\,V_{P} (S) \nonumber \\
E_{0} (S) &=& \phi_{0} (S) V_{P} (S) =
\frac{1}{\gamma}\,\phi_{0}(S') \, V_{P} (S) = \nonumber \\
&=& \phi_{0} (S') V_{P} (S') = E_{0} (S')
\end{eqnarray}

The interpretation of this result exhibits some interesting
features of the Lorentz transformation applied to electrodynamic
theory. It leaves the wave equations and their energy relations
intact if, and only if energy {\em quantities} are evaluated,
i.e. only in an integral evaluation of particle properties.
In this case it seems therefore justified, to consider
Special Relativity as a necessary adaptation of kinematic
variables to the theory of electrodynamics. The same does not
hold, though, for the intrinsic potentials of particle motion.
Since the
intrinsic potentials depend on the reference frame of evaluation
and increase by a transformation into a moving coordinate
system, the theory suggests that the infinity problems,
inherent to relativistic quantum fields, may be related to
the logical structure of QED.

\section{Conclusion}

We have shown that a new and {\em realistic}
approach to wave properties of moving particles allows to
reconstruct the framework of quantum theory and classical
electrodynamics from a common basis.

As it turns out, retrospectively, the theoretical framework
removes the three fundamental problems, which made a physical
interpretation (as opposed to the probability--interpretation
\cite{BOR26}) of intrinsic wave properties impossible.

\begin{itemize}
\item
The contradiction
between mechanical velocity of a particle and phase velocity of its  
wave is removed by the discovery of intrinsic potentials.
\item
The problem of intrinsic Coulomb interactions is removed,
because electrostatic interactions can be referred to an
exchange of photons. A particle in uniform motion is therefore
stable.
\item
The experimentally verified non--locality of micro physical
phenomena could be referred to measurements violating the
fundamental principles of quantum theory: they are therefore not
suitable to disprove any local and realistic framework.
\end{itemize}

From the viewpoint of electrodynamics the theory will reproduce any result
the current framework provides, because the basic relations are not
changed, merely interpreted in terms of electron and photon
propagation. Therefore an experiment, consistent with electrodynamics,
will also be consistent with the new theory.

In quantum theory, the only statements additional to the original
framework are beyond the level of experimental validity, defined
by the uncertainty relations. Since quantum theory cannot, in
principle, contain results beyond that level, every experiment
within the framework of quantum theory must necessarily be
reproduced. That applies to the results of wave mechanics, which
yields the eigenvalues of physical processes, as well as to the
results of operator calculations, if von Neumann's proof of the
equivalence of Schr\"odinger's equation and the commutation
relations is valid.

In quantum field theory, the same rule applies: the new theory only
states, what is, in quantum theory, no part of a measurement
result, therefore a contradiction cannot exist. If, on the other
hand, results in quantum electrodynamics exist, which are not yet
verified by the new theory, then it is rather a problem of
further development, but not of a contradiction with existing
measurements.

\begin{figure}
\epsfxsize=1.0\hsize
\epsfbox{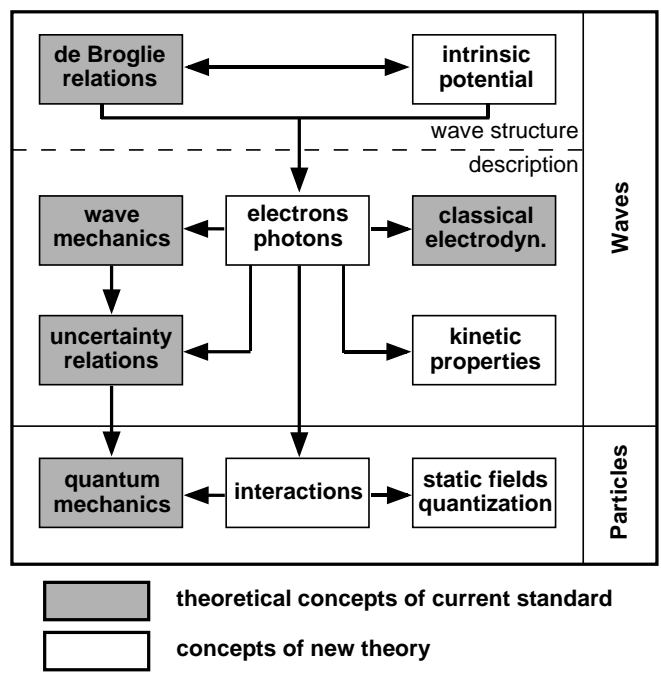}
\vspace{0.2 cm}
\caption{Logical structure of micro physics according to matter
wave theory. Classical electrodynamics and wave mechanics describe
intrinsic features of particles, quantization arises from
interactions.}
\label{fig001}
\end{figure}

Experimentally, the new theory cannot be disproved by existing
measurements, because its statements at once verify the existing
formulations and extend the framework of theoretical calculations.
The same does not hold, though, for the verification of existing
theoretical schemes of calculation by the new framework. Since the
theory extends far beyond the level of current concepts,
established theoretical results may well be subject to revision. The
theory requires that every valid solution of
a microphysical problem can be referred to physical properties of
particle waves. Essentially,
this is not limiting nor diminishing the validity of current
results, since it extends the framework of micro physics only
to regions, which so far have remained unconsidered.
The logical structure of the new framework and its relation to
existing theories is displayed in Fig. \ref{fig001}.

\section*{Acknowledgements}
Thanks are due to the {\em \"Osterreichische Forschungsgemeinschaft}
for generous financial support to attend the Athens conference. 




\begin{references}
\bibitem{BOR26}
Born M. {\it Z. Physik}\, {\bf 37}, 863 (1926)
\bibitem{COP59}
Bohr N. ''Discussion with Einstein on Epistemological Problems
in Atomic Physics'', in {\it A. Einstein: Philosopher--Scientist},
Schilpp P. A. (ed.), New York (1959)
\bibitem{BOH52}
Bohm D. {\it Phys. Rev.} \, {\bf 85}, 166; 180 (1952)
\bibitem{HEI27}
Heisenberg W. {\it Z. Physik} \, {\bf 43}, 172 (1927)
\bibitem{BEL64}
Bell J. S. {\it Physics} \, {\bf 1}, 195 (1964)
\bibitem{ASP82}
Aspect A., Dalibard J., and Roger G. \,{\it Phys. Rev. Lett.} \,
{\bf 49}, 1804 (1982)
\bibitem{PRI74}
George C., Prigogine I., and Rosenfeld L.
{\it Mat. Fys. Medd. Dan. Vid. Selsk.} \, {\bf 38} (12), 1 (1972)
\bibitem{EPR35}
Einstein A., Rosen N., and Podolsky B.
{\it Phys. Rev.} \, {\bf 47}, 180 (1935)
\bibitem{SCH92}
R\"oseberg U. in {\it Erwin Schr\"odinger's world view}, 
G\"otschl J. (ed.), Dordrecht (1992)
\bibitem{BRO57}
de Broglie L. in his foreword to \cite{BOH57}
\bibitem{BOH57}
Bohm D. {\it Causality and Chance in Modern Physics},
Princeton (1957) 
\bibitem{BEL82}
Bell J. S. {\it Found. Phys.} \, {\bf 12}, 989 (1982)
\bibitem{SCHW94}
Schweber S. {\it QED and the Men Who Made It},
Princeton (1994)
\bibitem{DIR84}
Dirac P. A. M. {\it Eur. J. Phys.} \, {\bf 5}, 65 (1984)
\bibitem{BRO26}
de Broglie L. {\it C.R. Acad. Sci. Paris} \, {\bf 183}, 447 (1926);
{\bf 185}, 580 (1927)
\bibitem{BRO60}
de Broglie L. {\it Nonlinear Wave Mechanics}, Amsterdam (1960)
\bibitem{BEL66}
Bell J. S. {\it Rev. Mod. Phys.} \, {\bf 38}, 447 (1966)
\bibitem{HOL93}
Holland P. R. {\it The Quantum Theory of Motion}, Cambridge (1993)
\bibitem{SCH26}
Schr\"odinger E. {\it Ann. Physik} \, {\bf 79}, 361; 489 (1926)
\bibitem{HEI25}
Heisenberg W. {\it Z. Physik} \, {\bf 33}, 879 (1925);
Born M., Heisenberg W., and Jordan P. {\it Z. Physik} \, {\bf 35},
357 (1926)
\bibitem{DAV27}
Davisson C. and Germer L. H. {\it Phys. Rev.} \, {\bf 30}, 705 (1927)
\bibitem{BRO25}
de Broglie L. {\it Ann. Phys.} \, {\bf 3}, 22 (1925)
\bibitem{PLA01}
Planck M. {\it Ann. Physik} \, {\bf 4}, 553 (1901)
\bibitem{COH77}
Cohen--Tannoudji C., Diu B., and Laloe F. 
{\it Quantum Mechanics}, New York (1977)
\bibitem{HOF97B}
Hofer W. A. {\it Measurement processes in quantum physics:
a new theory of measurements in terms of statistical ensembles},
to be published
\bibitem{JAC84}
Jackson J. {\it Classical Electrodynamics}, New York (1984)
\bibitem{EIN06}
Einstein A. {\it Ann Physik} \, {\bf 20}, 627 (1906) 
\bibitem{REN53}
Renninger M. {\it Z. Physik} \, {\bf 136}, 251 (1953)
\bibitem{UHL25}
Uhlenbeck G. E. and Goudsmit S. A.
{\it Naturwiss.} \, {\bf 13}, 953 (1925);
{\it Nature}\,{\bf 117}, 264 (1925)
\bibitem{ASP81}
Aspect A. PhD thesis, Universite de Paris--Sud, Centre d' Orsay
(1983)
\end{references}
\end{document}